\documentclass[aps,pre,twocolumn,letterpaper,10pt,showpacs,superscriptaddress]{revtex4-1}
\usepackage[english]{babel}
\usepackage{amsmath, mathtools,amssymb} 
\usepackage[usenames,dvipsnames,svgnames,table]{xcolor}
\usepackage[normalem]{ulem}
\usepackage{bm}
\usepackage{pifont}

\newcommand{\avg}[1]{\left< #1 \right>} 
\newcommand{\bracket}[1]{\left\langle #1\right\rangle}
\newcommand{\beeq}[1] {\begin{equation}\begin{split}#1\end{split}\end{equation}}

\newcommand{\LS}{Szil\'{a}rd}

\usepackage{hyperref}
\hypersetup{colorlinks=true, linkcolor=NavyBlue, citecolor=PineGreen,urlcolor=cyan}

\usepackage{xcolor}

\begin{document}
\title{Optimal work extraction and mutual information in a generalized \LS\ engine}

\author{Juyong Song}
\affiliation{Samsung Research, Samsung Electronics Co., Ltd., Seoul, 06765, Korea}

\author{Susanne Still}
\affiliation{Department of Information and Computer Sciences, and Department of Physics and Astronomy, University of Hawai'i at M\=anoa, Honolulu, 96822, Hawaii, USA }

\author{Rafael D\'iaz Hern\'andez Rojas} \email{rafael.diazhernandezrojas@uniroma1.it}
\affiliation{Dipartimento di Fisica,  Universit\`{a} di Roma ``La Sapienza'', P.le A. Moro 5, I-00185 Rome, Italy}

\author{Isaac P\'erez Castillo}
\affiliation{Departamento de F\'isica, Universidad Aut\'onoma Metropolitana-Iztapalapa, San Rafael Atlixco 186, Ciudad de M\'exico 09340, Mexico}

\author{Matteo Marsili} 
\affiliation{The Abdus Salam International Centre for Theoretical Physics (ICTP), Trieste, 34151, Italy}

\begin{abstract}
A 1929 Gedankenexperiment proposed by \LS, often referred to as ``\LS's engine", has served as a foundation for computing fundamental thermodynamic bounds to information processing. While \LS's original box could be partitioned into two halves and contains one gas molecule, we calculate here the maximal average work that can be extracted in a system with $N$ particles and $q$ partitions, given an observer which counts the molecules in each partition, and given a work extraction mechanism that is limited to pressure equalization. We find that the average extracted work is proportional to the mutual information between the one-particle position and the vector containing the counts of how many particles are in each partition. We optimize this quantity over the initial locations of the dividing walls, and find that there exists a critical number of particles $N^{\star}(q)$ below which the extracted work is maximized by a symmetric configuration of the $q$ partitions, and above which the optimal partitioning is asymmetric. Overall, the average extracted work is maximized for a number of particles $\hat{N}(q)<N^{\star}(q)$, with a symmetric partition. We calculate asymptotic values for $N\rightarrow \infty$. 
\end{abstract}

\maketitle

\section{Introduction}
The thought experiment known as ``Maxwell's Demon''~\cite{maxwell1871theory} 
addressed the issue that the Second Law of thermodynamics is statistical in nature. An ideal gas at temperature $T$ is enclosed in an isolated container divided into two equal parts by a fixed wall with a trap door operated by some sentient being, later called a ``demon'', who will open the door to incoming particles, sorting them by velocity. This process would result in a temperature gradient that could be used to obtain work from the system. Maxwell's idea started an ongoing debate, to which \LS\ contributed significantly with a model that circumvents the necessity of a sentient being, replacing it by a simple mechanism which, importantly, retained the main feature of the ``demon", namely that of having a memory. 

\LS's engine consists of a single particle gas within a container divided into two equal partitions separated by a movable, frictionless wall \cite{szilard-german}.  When the container is put into contact with a single thermal reservoir at temperature $T$, the movable wall may then be used to extract work (\textit{e.g.} by lifting a weight), when moved towards the empty side of the box, as the particle transfers kinetic energy in successive elastic collisions. Being able to do this requires knowledge of which side is empty at the beginning of work extraction, to remain present throughout, i.e., it requires a memory. When operated cyclically, the average extractable work is compensated by the average amount of work that has to be done to run the memory. An adiabatic, isothermal volume expansion yields work $W_{\text{ext}}= k_B T \ln\left({V \over V/2}\right) = k_B T \ln 2$, which corresponds to $k_B T$ times the mutual information captured about the coarse grained particle location. This idea has served as a foundation not only for computing fundamental thermodynamic bounds for information processing (e.g. \cite{landauer1961irreversibility, bennett1982thermodynamics, sagawa2012thermodynamics, parrondo2015thermodynamics, CB}), but also for concretely demonstrating how information can be turned into work, and vice versa. Recently, interest in these issues has spiked with increased experimental capabilities \cite{sagawa2009minimal, berut2012expLandauer, mandal2012work, SagawaUedaFeedback12, mandal2014, koski2014experimental, exp-landauer2014, martinez2016brownian, hong2016experimental, gavrilov2016erasure, gavrilov2017direct, lathouwers2017memory, kumar2018nanoscale, paneru2018lossless, admon2018experimental,wolpertbook2019}. Different variations of the standard \LS\ engine have been discussed, including generalisations to $N$-particles systems \cite{kim2011quantum,kim2011information} and non-ideal (classical or quantum) particles \cite{Horowitz_2011,bengtsson2018quantum}.

Here, we study how much work can be extracted, on average, when a \LS\ engine, operated quasi-statically, contains an ideal gas with $N$ particles, and when $q$ partitions can be created in the box. We assume that the observer counts and memorizes how many particles fall into each partition, and then exploit the isothermal expansion of the ideal gas in the different compartments to extract work, as in the original \LS\ box \cite{szilard-german}. 
We show in Sec. \ref{sec2} that the average extracted work is proportional to the mutual information retained in memory about the location of a single particle, not of the location vector of the ensemble. The latter information controls the minimal cost for memorizing the counts, whereby their difference controls a lower bound on dissipation of the engine (when run cyclically). We calculate how much average work can maximally be extracted when the choice of where to place the movable walls before the measurement is optimized, for fixed $N$ and $q$. To build intuition, Sec. \ref{sec3} treats the case with only one movable wall, $q=2$, and confirms agreement with previous work. The general case is then treated in Sec. \ref{sec4}.

\section{Mutual information and work} 
\label{sec2}
\begin{figure}[h]
\centering
\includegraphics[width=\linewidth]{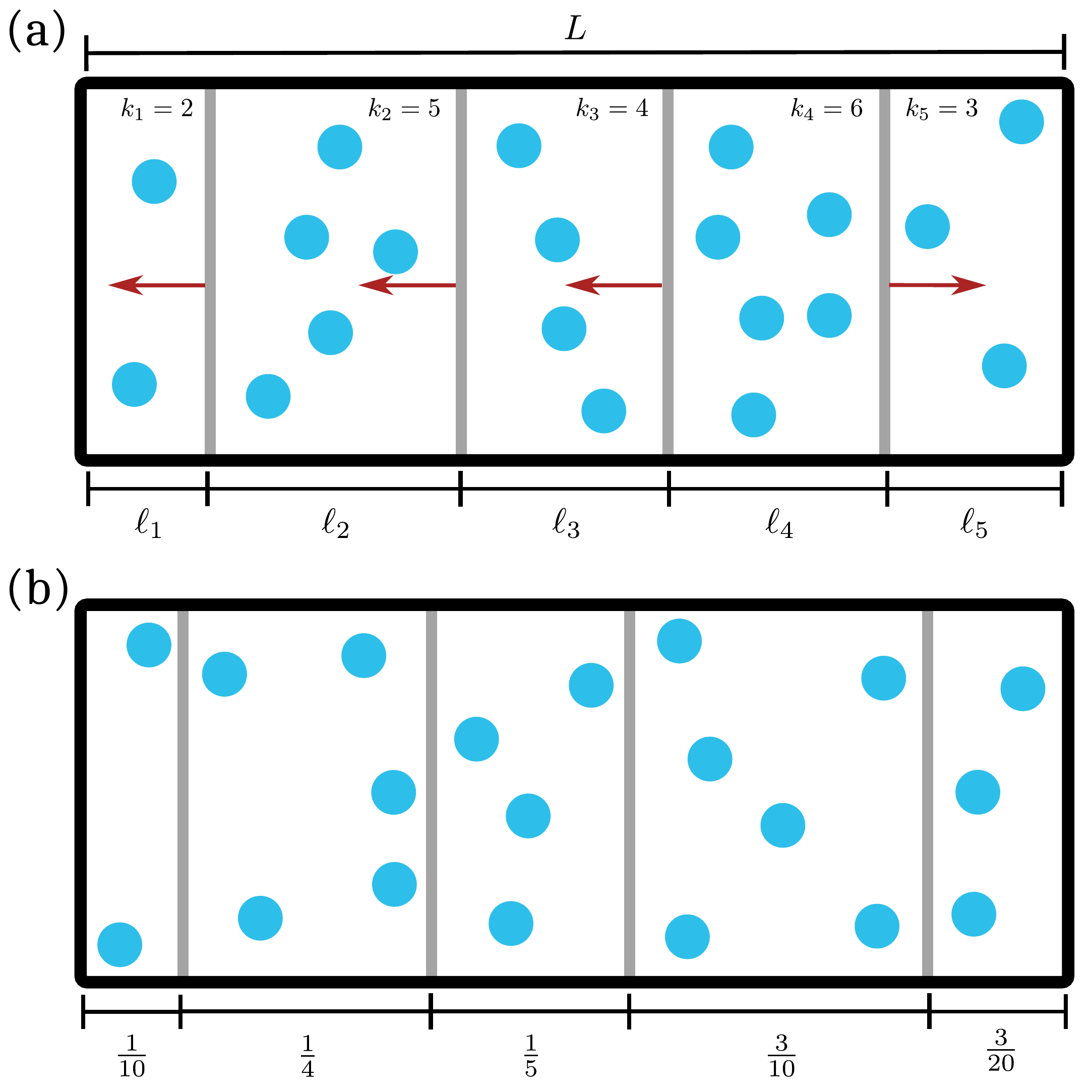}
\caption{Schematic representation of a Szilard engine with $q=5$ partitions and $N=20$ particles. Panel (a) depicts the initial state, where the length of each partition is given by $\ell_i$ and the red arrows indicate in which direction each wall is going to move as the engine performs work. The final state, which is assumed to be reached quasi-statically is shown in panel (b), where the length of each partition is given by $\ell_i^{(\rm eq)} = k_iL/N$ and thus the pressure is the same in all of them. The values indicated in the lower panel corresponds to the counts obtained by setting $L=1$.
}
\label{fig:diagram}
\end{figure}

We consider a \LS\ engine, generalized to $N$ particles inside a container of longitudinal size $L$ and transverse unit area. 
The gas in the container is coupled to a thermal reservoir, and is in thermal equilibrium at the beginning of each cycle. The mass of the walls is assumed to be much larger than the mass of the particles (which is set to unity). The engine is run cyclically as follows:
\begin{enumerate}
\item Preparation step (assumed not to require work): Insert walls which divide the container along the longitudinal axis into $q$ partitions with lengths \mbox{$\bm{\ell}= (\ell_1, \ldots, \ell_q)$}, see Fig.~\ref{fig:diagram}a. 
\item Measurement and data representation step: the observer has {\it a priori} knowledge of the experimental setup, including 
$\bm{\ell}$, and is provided with a snapshot of the $N$ particles' x-positions, denoted by $\bm{x}=(x_1, \ldots, x_N)$. Using this measurement, the observer commits the counts of how many particles reside in each partition to memory, denoted by $\bm{k}=(k_1, \ldots, k_N)$.
\item Work extraction step: using the information committed to memory, weights are attached in such a way that work is extracted while the walls move quasi-statically until the pressure is equalized across the container. The observer knows the number density in each partition, $k_i/\ell_i$, and thus the local pressure, $P_i=k_B T k_i/\ell_i$ \footnote{The volume of partition $i$ is $\ell _i$, because the container has transverse unit area.}. This enables the observer to determine how much each partition will be moved by the expansion of the gas, and in which direction. In equilibrium, the pressure is $P_i^{(\rm eq)}=N k_BT/L$ in all partitions, which implies that the lengths of the partitions after work extraction are $\ell_i^{(\rm eq)}=k_i L/N$. 
\item Return to beginning: Walls are pulled out (assumed not to require work).
\end{enumerate}

We know that the total information captured by the observer's memory, about the available data determines the energetic cost of the memory, while only the relevant part of this information determines the potential energetic gain \cite{CB}. To asses the engine's overall efficiency, we need to calculate both aspects. 

In the setup we consider here, available data are the x-positions of the ensemble, $\bm{x}$, and work extraction can happen only via movement of the walls towards pressure equalization. Since the particles of an ideal gas are non-interacting, only information about the locations of the $N$ individual particles matters, none of the correlations between particle positions make any difference to calculating the density, and thus the pressure, in the different partitions. 

But the counts $\bm{k}$ do contain information about the ensemble, because knowledge of ${k_i}$ constraints the possible counts in the other partitions, $k_{j\neq i}$. 
Counting is thus intrinsically wasteful in this situation, because it captures some irrelevant information. Overall engine dissipation is proportional to irrelevant information \cite{CB} when the observer is viewed as a part of the information engine and the observer's energetic costs are taken into account. 

We thus have a quantitative expectation of the costs and gains associated with the observer's memory, given the physical constraints imposed by our specific work extraction protocol. To make that precise, let $\bm{K}$ denote a random variable with realizations $\bm{k} \in \mathcal{K}$. This, together with $p(\bm{k}|\bm{x})$, characterizes the memory we have chosen here: ensemble values of all N $x$-positions are mapped onto memory states, which are specified by the count vector $\bm{k}$. 
We are using the following standard notational shortcuts: vectors are bold face symbols, entropy functionals are written as $H[X] = -\langle \ln{p(x)} \rangle_{p(x)}$, and \mbox{$H[\bm{X}] = -\langle \ln{p(\bm{x})} \rangle_{p(\bm{x})} = -\langle \ln{p(x_i, \dots, x_N)} \rangle_{p(x_i, \dots, x_N)}$}, where $\langle \cdot \rangle_{p}$ denotes the average. Conditional entropy and information are written accordingly \cite{CoverThomas}. For readers unfamiliar with information theory, all terms and calculations pertaining to this section are written out in detail in Appendix \ref{A}.

While the counts committed to memory capture information in the amount of $I[\bm{X}, \bm{K}]$, the given work extraction protocol should not allow the observer to recover that much work. We expect that the observer can use only the total information captured about individual particle locations, which is
$\sum_{i=1}^N I[X_i,  \bm{K}] = N I[X,\bm{K}]$, because the particles are identical. 

\subsection{Extractable work and relevant information}
\label{W-I}

The work extracted \cite{callen1998thermodynamics} when all partitions change from $\ell_i$ to $ k_i L/N$, is
\beeq{
W(\bm{k})&=\sum_{i=1}^q\int_{\ell_i}^{L\frac{k_i}{N}} d V_i P_i= k_B T \sum_{i=1}^q\int_{\ell_i}^{L\frac{k_i}{N}} d V_i k_i/V_i \\
&= k_BT\sum_{i=1}^q k_i \ln \frac{k_iL}{N\ell_i}\,,
\label{eq:work-extr}
}
The expected extracted work \footnote{Averages are denoted by $\bracket{\cdot}$}, $\bracket{W}$, then results from averaging $W(\bm{k})$ over the distribution $P(\bm{k})$ of measurement vectors given by the multinomial distribution 
\beeq{
P(\bm{k}) = \frac{N!}{k_1! \cdots k_q!} \, p_1^{k_1} \cdots\, p_q^{k_q}, \label{multin dist}
}
where $p_i = \ell_i/L$ is the probability of finding any one particle in partition $i$.

We now calculate the relevant information captured by the counts:
\beeq{
I[X ,\bm{K}] 
&= \sum_{\bm{k}} P(\bm{k})\int {\rm d}x P({x}|\bm{k}) \ln{\frac{P({x}|\bm{k})}{P(x)}} ~. 
}

First, note that
the probability of finding a single particle in any location along the x-axis is $P(x) = 1/L$.  Second, within each partition, $j$, the overall probability of finding a particle, given the count vector $\bm{k}$, is $k_j/N$.
The probability of finding a single particle in position $x$ within partition $j$ is uniform over the length of the partition, $\ell_j$. Therefore, the probability of finding a particle in position $x$, given the counts $\bm{k}$ is:
	
\beeq{
P(x|\bm{k}) = \frac{k_i}{N \ell_i},\quad \sum_{j=0}^{i-1} \ell_j \leq x < \sum_{j=1}^{i} \ell_j\, ,
\label{eq:Pxgk}
}
for $i=1,\ldots,q$, using the convention $\ell_0=0$. 

Putting everything together lets us compute how much information the counts contain about the location of a single particle:
\beeq{
I[X ,\bm{K}] 
&=\left\langle  \sum_{i=1}^q \frac{k_i}{N} \ln \frac{k_i L}{N \ell_i} \right\rangle_{P(\bm{k})} \, .
\label{eq:a}
}

We thus arrive at our main result: combining Eqs. (\ref{eq:work-extr}) and (\ref{eq:a}), tells us that the average extracted work is proportional to $N$ times the single particle location information captured by the counts,
\begin{equation}
\label{mainresult}
\bracket{W}= k_B T NI[X,\bm{K}]. 
\end{equation}

\subsection{Irrelevant information retained in memory}
To physically run the memory that keeps the counts of particles in each partition, heat in the amount of at least $k_B T I[\bm{X},\bm{K}]$ joules has to be dissipated per cycle, on average (e.g. \cite{parrondo2015thermodynamics, CB}). But the memory can be used to extract, on average, work up to only $k_B T NI[X,\bm{K}]$ joules per cycle. The lower limit on overall average dissipation per cycle is the difference, which is proportional to the irrelevant information retained in memory \cite{CB}. Here, the counts keep irrelevant information in the amount of
\begin{eqnarray}
I_{\rm irrel} &=& I[\bm{X},  \bm{K}] - N I[X, \bm{K}] \label{Iirrel-1}\\
&=& I[\bm{X} | \bm{K}] \geq 0 \label{Iirrel-2}~,
\end{eqnarray}
equal to the conditional multi information, 
\begin{equation}
I[\bm{X} | \bm{K}] = \left\langle \ln{\left[ {P(x_1, \dots, x_N | k_1, \dots, k_q) \over \prod_i^{N} P(x_i | k_1, \dots, k_q)} \right]} \right\rangle_{P(\bm{x},\bm{k})}~, \label{Iirrel-3}
\end{equation}
a non-negative quantity.
To get from line (\ref{Iirrel-1}) to line (\ref{Iirrel-2}), note that $I[\bm{X}, \bm{K}] - \sum_{i=1}^N I[X_i,  \bm{K}] = I[\bm{X} | \bm{K}] - I[\bm{X}]$. But the multi information $I[\bm{X}] = 0$ is zero here, because the particles are non-interacting.

The mapping from particle locations to the vector of counts characterizes the memory accessible to the information engine. Counting is a deterministic mapping, which means that the conditional entropy $H[\bm{K}|\bm{X}]$ is zero, since the counts are completely determined by the locations (unless measurement errors are taken into account). Therefore, we have $I[\bm{X},\bm{K}] = H[\bm{K}]$. The irrelevant information retained in the counts is thus $I_{\rm irrel}^{\rm count} =$
\begin{eqnarray}
H[\bm{K}] - N I[X,\bm{K}] 
=  \ln\left[\frac{N^N}{N!}\right]  - \sum_{i=1}^q \left\langle  \ln \left[ \frac{k_i^{k_i}}{k_i!} \right] \right\rangle_{p(k_i)}
\end{eqnarray}
We use Stirling's approximation, $n!\cong n^n e^{-n}\sqrt{2\pi n}$, to estimate for very large $N$, and for equidistant partitions:
\begin{eqnarray}
I_{\rm irrel}^{\rm count} &\cong&
{1\over 2} \Bigl( (q-1) \ln(2\pi N) -q\ln(q)\Bigr)
\end{eqnarray}


\subsection{Discussion}
Better memories could perhaps do better, as could different work extraction mechanisms. The problem of finding an optimal memory was discussed for the case of generalized, partially observable information engines in \cite{CB}. The case where particles have a weak, short range, repulsive potential is studied in \cite{Horowitz_2011}. Furthermore, it is obvious that if the particles can initially be prepared in a non-equilibrium state, then a larger amount of work can be extracted from a cycle of the Szil\'ard engine, a point made in \cite{Touzo_2020}. However, none of these variations on the original \LS\ scheme are the focus of this paper. 

Instead, here we explore the consequences of the choices made in our specific setup: a naive, counting, observer, and a work extraction mechanism based solely on pressure equalization. These choices constitute a {\em direct} extension of the \LS\ engine to $N$ particles and $q$ partitions, without the introduction of additional variations or optimizations. 
In this straightforward extension of the \LS\ engine to $N$ particles and $q$ partitions there is then only one remaining choice, namely the initial placement of the dividers. Conceptually, we could ascribe the task of inserting the partitions to the observer, which would also justify the fact that the observer knows $\bm{\ell}$. 

We will now compute those partition locations $\bm{\ell}$ that maximize relevant information, and hence maximize extractable work (Sec.~\ref{W-I}). This tells us how to design the best naive $N$ particle $q$ partition classical extension of \LS's engine. 

\section{Maximizing work extraction by positioning divider partitions}
\label{Max-l}

We now ask: given a system of $N$ particles and $q$ partitions, does there exist a partitioning that is optimal in the sense that it maximizes the average extractable work? That means, we seek to find a location vector $\hat{\bm{\ell}}$ that maximizes $I[X ,\bm{K}]$.\\

\subsection{{Maximum} work extraction for a single movable wall}

\label{sec3}
To build intuition, let us start with the case of one movable wall, i.e., two partitions.  Let $p$ denote the probability of finding a particle in the left partition of longitudinal size $\ell$, which is $p=\ell/L=\bracket{k}/N$. The probability of observing $k$ particles out of the $N$ possible in the left partition is then:
\beeq{
P(k)=\binom{N}{k}p^k (1-p)^{N-k}\,.
}
The conditional probability $P(x|k)$ is:
\beeq{
P(x|k) = \begin{cases}
k\, / \, {N \ell}, & 0 \leq x \leq \ell\\
(N-k)\, / \, {N (L-\ell)}, & \ell < x \leq L
\end{cases}\,,
}
and the marginal probability is $P(x)=1/L$. 
This yields the mutual information between the measurement $k$ and the position of any single particle:
\beeq{
I[X, K] \!=\! \sum_{k} \!P(k) \Bigg[ {k\over N} \ln{\frac{k L}{N \ell}} + {N-k\over N} \ln{\frac{(N-k) L}{N (L-\ell)}} \Bigg]\,,
\label{eq:1}
}
which is in agreement with earlier work \cite{kim2011quantum,kim2011information}.

For one particle, a quick and intuitive calculation shows that the maximal value of $I[X, K] $ is attained by placing the wall in the middle. But this is not always the case for any number of particles, $N$.
\begin{figure} 
\centering
\includegraphics[width=8.5cm,height=5.5cm]{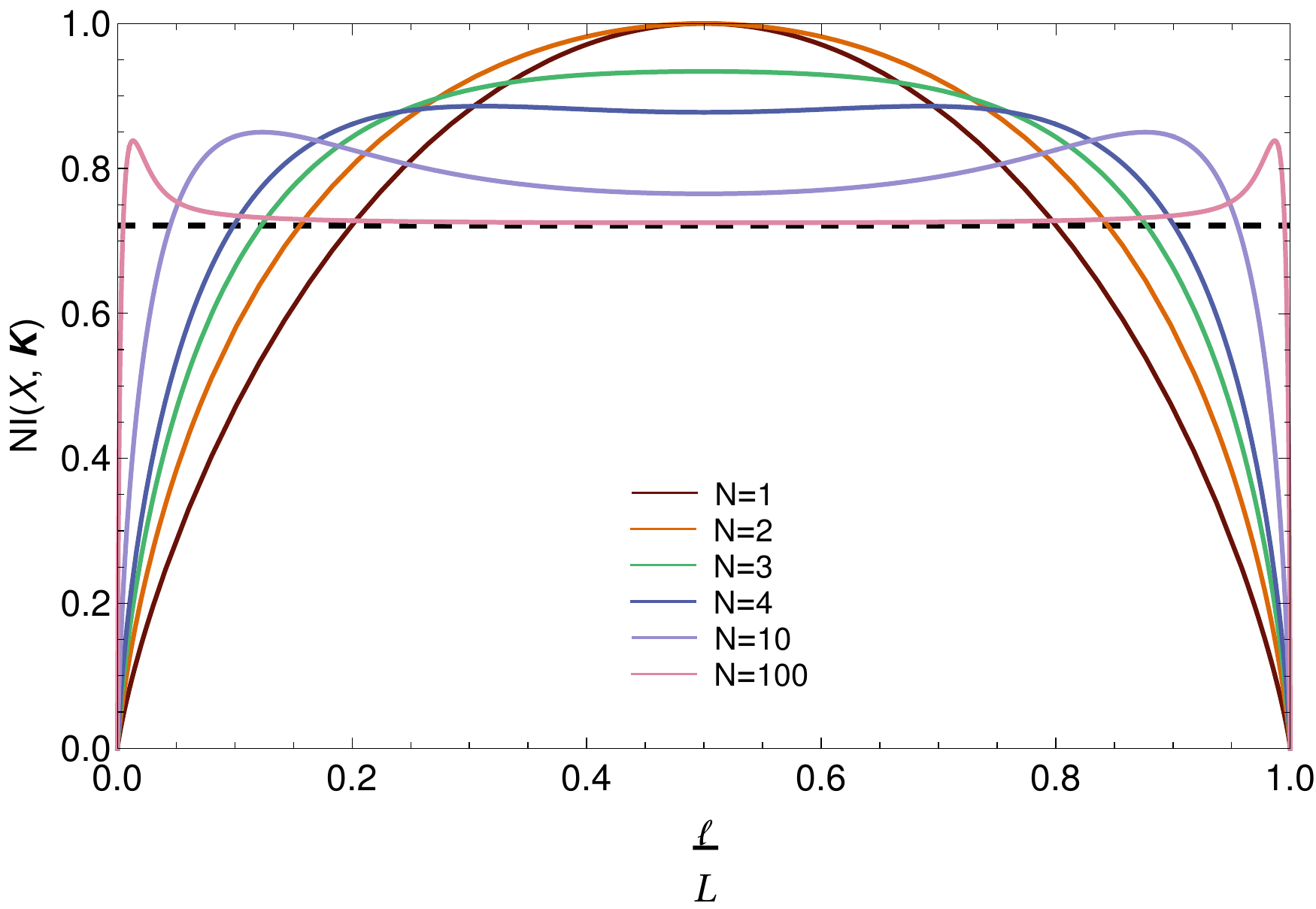}
\caption{The mutual information $N\, I[X,\bm{K}]$ as a function of the  position of the wall $p=\ell/L$, for a \LS's engine containing $N$ particles. The different curves correspond to different values of $N$, while the dotted line is the asymptotic limit  ${1\over 2\ln (2)} \simeq 0.7213$ bits . }
\label{fig1}
\end{figure}
To illustrate this, we plot $N I[X, K] $ against $p=\ell/L$ in Fig. \ref{fig1}, for various values of $N$. Only for $N\leq 3$ does the optimal position of the movable wall, $\hat{\ell}$, correspond to halving the volume ($\hat{\ell}=L/2$). For larger numbers of particles, the optimal partition is given by an asymmetric configuration of the partitions: the symmetric solution with the wall in the middle becomes a local minimum, and two maxima appear at $\hat{\ell}$ and $1-\hat{\ell}$. This agrees with what was reported in \cite{pal2016role}. To understand the mechanism behind the appearance of this asymmetric solution, we can perform an asymptotic expansion of $NI[X,K]$ for large $N$. Introducing the average count $n:=\bracket{k}=N~\ell/L$, and the random variable $\Delta$, such that $k=n+\Delta$, we first write the mutual information as:
\beeq{
N I[X,K]&=\sum_{k}P(k)\left[k\ln \frac{kL}{\ell N}+ (N-k)\ln\frac{(N-k)L}{N(L-\ell)}\right]\\
&=\sum_{k}P(k)\Bigg[\left(n + \Delta\right) \ln\left(1 + \frac{\Delta}{n}\right)\\
&+ \left(N-n-\Delta\right) \ln\left(1-\frac{\Delta}{N-n}\right)\Bigg]\,,
}
and then perform a Taylor expansion for small values of $\Delta$. Recalling the following statistical properties of the binomial distribution, $\avg{\Delta}=0$, $\avg{\Delta^2} = N p\, (1-p) $ , $\avg{\Delta^3} = N p (1-p) (1-2p)$ and $\avg{\Delta^4} = 3 N(N-2) p^2 (1-p)^2 + N p (1-p)$, we obtain
\beeq{
N I[X, K] = \frac{1}{2} + \frac{1}{4N} + \frac{(1-2p)^2}{12 N p (1-p)} + \mathcal{O}\left({N^{-2}}\right)\,.
\label{explargeN}
}
Apart from the first term, all orders go to zero in the limit of an infinite number of particles. For finite $N$, however, the second term gives the leading order on the decaying value for the symmetric partition, while the third one is responsible of making the symmetric partition become a local minimum, because, as $p$ deviates from $1/2$, this term increases. The same asymptotic expansion can be carried out for the general case of $q$ partitions, which we will calculate later. 

For a more quantitative analysis, we use an integral representation of the natural logarithm,
\begin{equation}
\ln z = \int_{0}^{\infty} \frac{du}{u} \left( e^{-u} - e^{-zu} \right)\,,
\label{Eq::log_int}
\end{equation}
to rewrite Eq.~\eqref{eq:a} as
\begin{eqnarray}
NI[X,\bm{K}]&=\sum_{i=1}^q\sum_{k_i}P(k_i) & k_i\ln\left(\frac{k_i }{\bracket{k_i}}\right)\\
&=\sum_{i=1}^q\int_0^\infty\frac{du}{u}\Bigg[ & \bracket{k_i\left(e^{-u} - e^{-k_iu}\right)}\\
&& -\bracket{k_i}\left(e^{-u} - e^{-\bracket{k_i}u}\right)\Bigg]~. \notag
\end{eqnarray}
Noticing that
\beeq{
\bracket{e^{-k_i u} }&= \sum_{k_i=0}^{N} {N \choose k_i} (p_i e^{-u})^{k_i} (1-p_i)^{N-k_i}\\ 
&= \left( 1-p_i(1 - e^{-u} ) \right)^N\,,
}
we obtain
\beeq{
N I[X,\bm{K}]	&=\sum_{i=1}^{q}\mathcal{F}_N(n_i)\,,
\label{eq:I-F}	
}
with $n_i := \avg{k_i}$, and the function
$\mathcal{F}_N(x)$ is given by
\beeq{
\mathcal{F}_N (x) &= \int_0^\infty \!\frac{du}{u^2} \left\{ \left[ 1 - \frac{x}{N} (1-e^{-u})\right]^N  - e^{-x u}  \right\}\,.
\label{eq:3}
}
We can hence write Eq. \eqref{eq:1} as
\beeq{
N I[X, K] = \mathcal{F}_N (n) + \mathcal{F}_N (N-n)\,,
\label{eq:2}
}
where $n=\bracket{k}$ is the average count.

\begin{figure} 
\centering 
\includegraphics[width=8.5cm,height=5.5cm]{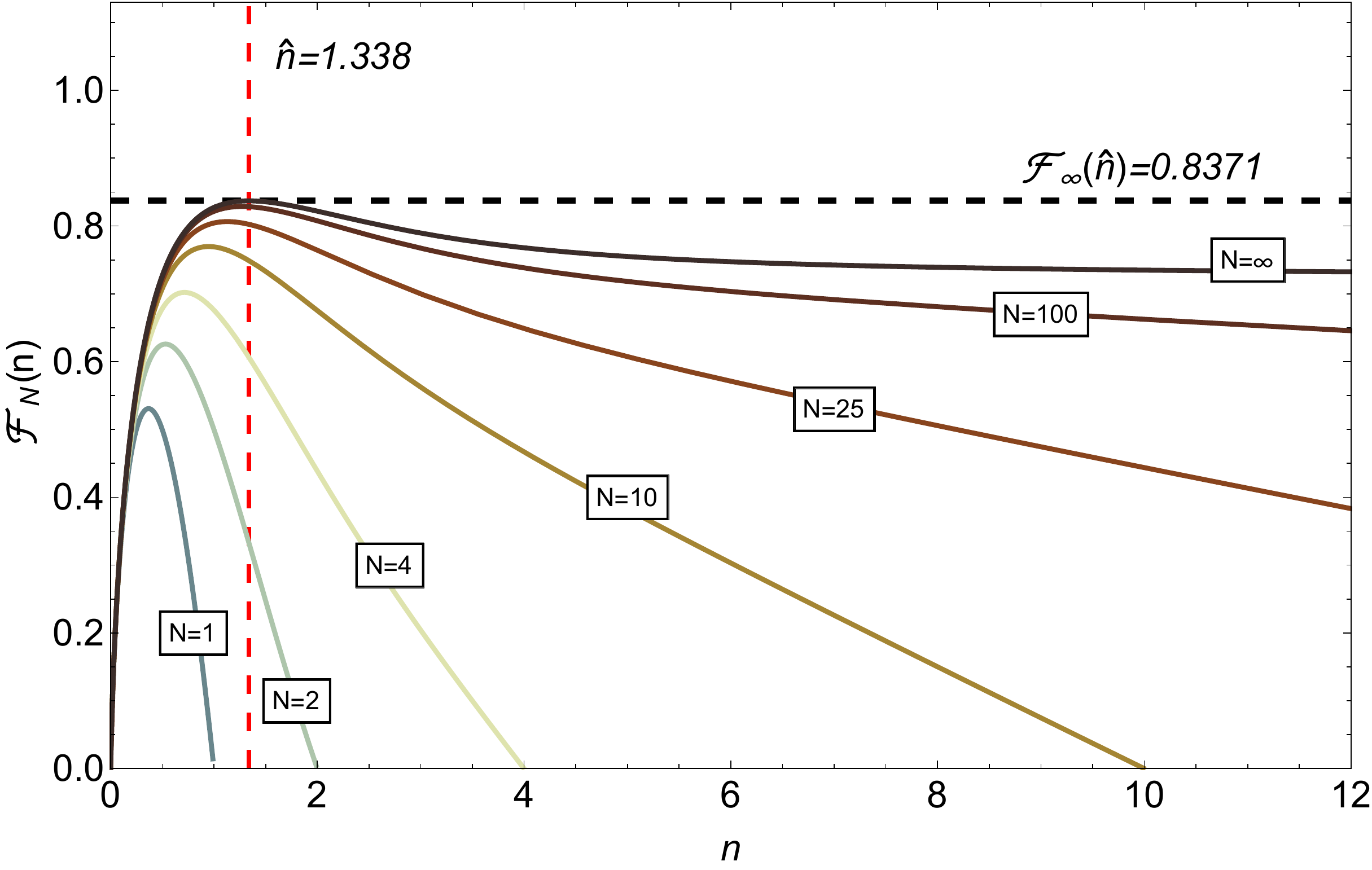}
\caption{$\mathcal{F}_N(n)/\ln(2)$ [in bits] as a function of $n$, plotted for various values of $N$. In the limit of an infinite number of particles the maximum of  $\mathcal{F}_N(n)$ is achieved at $\hat{n}\simeq 1.338$ and has a value of $\mathcal{F}_\infty(\hat{n})\simeq 0.8371$ bits.}
\label{fig2}
\end{figure}
Thus, maximizing $N I[X, K]$ with respect to the position $\ell$ of the wall for a fixed number of particles is equivalent to maximizing $\mathcal{F}_N(n) + \mathcal{F}_N (N-n)$ with respect to $n$. A plot of the function $\mathcal{F}_N(n)/ln(2)$ (scaled to be in units of bits) can be found in Fig.~\ref{fig2} for various values of $N$. The optimal value  $\hat{n}(N)$ which maximizes the mutual information must obey  $\mathcal{F}'_N(n)=\mathcal{F}_N'(N-n)$. 
For $N\leq 3$ one finds that $\hat{n}=\frac{N}{2}$, and for larger numbers of particles we have that the optimal solution is given by an asymmetric partition (consistent with Fig.~\ref{fig1}). For $N\rightarrow \infty$, either $\mathcal{F}_N (n)$ or $\mathcal{F}_N (N-n)$ becomes the dominant term in Eq.~\eqref{eq:2}, and one can write that
\beeq{
\lim_{N \to \infty} N I[X, K]&= \mathcal{F}_{\infty}(n) \\
&= \int_0^\infty \frac{du}{u^2} \left( e^{-n(1-e^{-u})} - e^{-nu} \right)\,.
\label{eq:3b}
}
The maximal value of the mutual information occurs at $\hat{n}(\infty)\simeq 1.338$ and is $\mathcal{F}_{\infty}(\hat{n})\simeq 0.8371$ bits. This is to be contrasted with a symmetric partition, which would give a lower value for the mutual information of ${1\over 2} \ln(2)\simeq 0.7213$ bits (see Fig.~\ref{fig1}).

The larger the number of particles, $N$, the closer to the edge of the box we have to insert the movable wall to maximize average work extraction, while the average work extracted by a wall in the middle goes to zero. This explains not only why, with a container filled with some regular gas ($N$ roughly between $10^{22}$ and $10^{23}$), zero work can be extracted, on average, by putting a wall in the middle, but also why there is no chance to extract macroscopic work by implementing the optimal partitioning, because the necessary distances become much too extreme to realize.

Optimizing the average extracted work (for $q=2$) also with respect to the number of particles, $N$, gives as the best choice either one or two particles, (with the wall in the middle), as can be appreciated from Figs.~\ref{fig1} and \ref{fig3}.

\subsection{Optimal work extraction with $q$ partitions}
\label{sec4}
For one particle, we can trivially insert as many partitions as our experimental setup allows, and measure to the same resolution, in order to get more work out of the information engine, but we equally have to spend more energy to run the memory. We have, for one particle, that the cost and the potential benefit of the memory are precisely equal, because $I[\bm{X},\bm{K}] = I[X ,\bm{K}]$ for $N=1$. Therefore, the overall bound on the engine's dissipation is unaffected. With one particle, a cyclically run \LS\ engine can, in principle, achieve zero dissipation. 

How much work can be extracted from a \LS\ box with $N$ particles and $q$ partitions? We can use Eq. \eqref{eq:I-F} to analyze this general case similarly to the $q=2$ case discussed above. Finding the optimal partition that maximizes the mutual information for fixed $N$ and $q$, \textit{i.e.}, $\hat{\bm{\ell}}=(\hat{\ell}_1,\ldots,\hat{\ell}_q):=\text{arg max}_{\bm{\ell}} \left( N I[X,\bm{X}]\right)$, together with the maximal value, $\hat{\mathcal{I}}_{q}(N) = \max_{\bm{\ell}}\left(  N I[X, \bm{K}]\right)$, then reduces to finding the number vector $\bm{n}=(n_1,\ldots,n_q)$ that maximizes Eq. \eqref{eq:I-F}. The number vector has to be normalized, $\sum_{i=1}^qn_i=N$, and to carry out the optimization, we introduce
\beeq{
\mathcal{I}^{(N)}_q(\bm{n},\lambda)=\sum_{i=1}^q\mathcal{F}_N(n_i)-\lambda\left(\sum_{i=1}^qn_i-N\right)\,,
}
which must be maximized with respect to $\{\bm{n},\lambda\}$, where $\lambda$ is a Lagrange multiplier. 
Then the optimal $\hat{\bm{n}}$ must obey
\beeq{
0&=\frac{\partial \mathcal{I}^{(N)}_q(\bm{n},\lambda)}{\partial n_i}\Big|_{\bm{n}=\hat{\bm{n}}}\,,\quad\quad i=1,\ldots,q\,,\\
0&=\frac{\partial \mathcal{I}^{(N)}_q(\bm{n},\lambda)}{\partial \lambda}\Big|_{\bm{n}=\hat{\bm{n}}}\,,
}
yielding:
\beeq{
\lambda&=\mathcal{F}_N^\prime(\hat{n}_i)\,,\quad\quad i=1,\ldots,q\,,\\
N&=\sum_{i=1}^q\hat{n}_i\,.
}
We must, moreover, be sure that 
\beeq{
\sum_{i,j=1}^q\frac{\partial^2 \mathcal{I}^{(N)}_q(\bm{n},\lambda)}{\partial n_i\partial n_j}\Big|_{\bm{n}=\hat{\bm{n}}}\delta n_i\delta n_j<0\,,
}
with $\delta n_i=n_i-\hat{n}_i$, or equivallently,
\beeq{
\sum_{i=1}^q \mathcal{F}_N''(\hat n_i)(\delta n_i)^2<0\,.
\label{eq:cond}
}
with $\sum_{i=1}^q\delta n_i=0$.

\begin{figure}
\centering
\includegraphics[width=8.5cm,height=5.5cm]{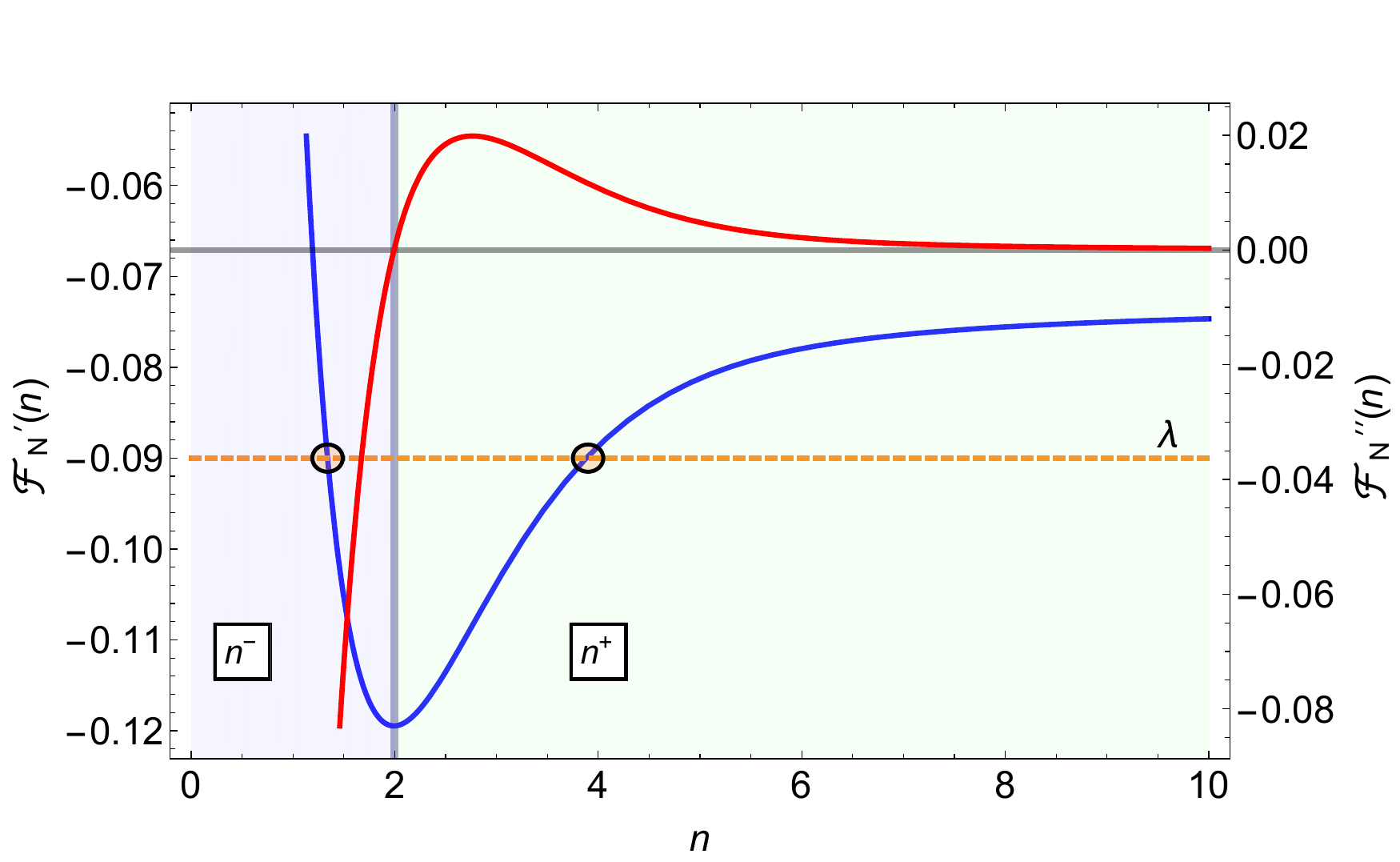}
\caption{Plot of  $\mathcal{F}_N'(n)$ (solid blue line) and $\mathcal{F}_N''(n)$ (solid red line) as a function of $n$, for $N=10$. As we can see here, for a given value of $\lambda$ (dashed orange line), there are two solutions of $\mathcal{F}_N'(n)=\lambda$ (represented here with black circles). These, denoted as $n^{-}$ and $n^{+}$ are such that $\mathcal{F}_N''(n^{-})<0$ and  $\mathcal{F}_N''(n^{+})>0$. }
\label{fig:app1}
\end{figure}

To understand the solutions to this system of equations, we plot $\mathcal{F}_N'(n)$ and $\mathcal{F}_N''(n)$ for different values of $N$ and $q$ in Fig. \ref{fig:app1}. The symmetric solution $n_i=N/q$ is stable as long as $\mathcal{F}_N''(N/q)<0$ (red curve), but beyond that point unstable, and we need to investigate asymmetric solutions. For a given value of $N$ and $q$, the equation $\lambda=\mathcal{F}_N'(\hat{n}_i)$ has two solutions $\hat{n}_i=n^{-}$ and $n^{+}$ with $n^{-}<n^{+}$. Since the Lagrange multiplier imposes a global constraint, this implies that, regardless of the order of the indices labeling the partitions, a general solution may correspond to  having $q^{-}$ partitions with solution $n^{-}$ and $q^{+}$ partitions with $n^{+}$, such that $q=q^{-}+q^{+}$ and $N=q^{-}n^{-}+q^{+}n^{+}$. Importantly, the solution $n^{-}$ is such that   $\mathcal{F}''_N(n^{-})\leq 0$, while for $n^{+}$ we have instead that $\mathcal{F}''_N(n^{+})\geq 0$ (see solid red line in Fig. \ref{fig:app1}). Therefore, the stability condition \eqref{eq:cond} reads
\beeq{
 \left|\mathcal{F}''_N(n^{+})\right|\sum_{\ell=1}^{q_+}(\delta n^{+}_\ell)^2-\left|\mathcal{F}''_N(n^{-})\right|\sum_{\ell=1}^{q_-}(\delta n^{-}_\ell)^2<0\,,
  \label{eq:cond2} 
}
with
\beeq{
\sum_{\ell=1}^{q_{+}}\delta n^{+}_\ell+\sum_{\ell=1}^{q_{-}}\delta n^{-}_\ell=0\,.
  \label{eq:cond3} 
}
This automatically implies that $q^{+} \leq 1$, because if $q^{+}$ was 2 or larger, the inequality (\ref{eq:cond2}) could be violated by the choice $\delta n^{-}_\ell=0$ for all $\ell=1,\ldots q^{-}$. This means that the optimal asymmetric partition corresponds to having one large partition and $q-1$ small and equal partitions. 

\begin{figure} 
\centering 
\includegraphics[width=8.5cm,height=5.5cm]{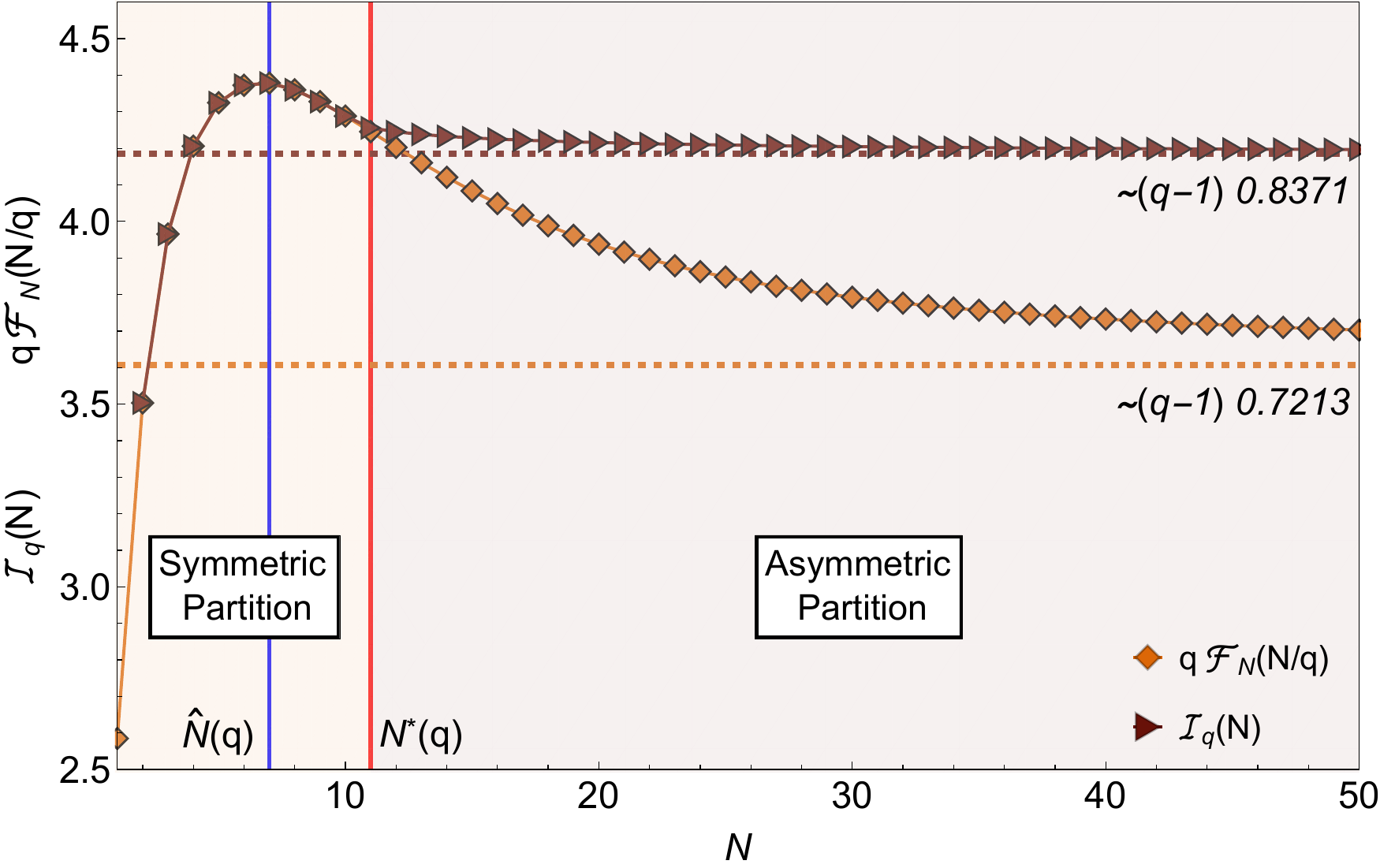}\\
\includegraphics[width=8.5cm,height=5.5cm]{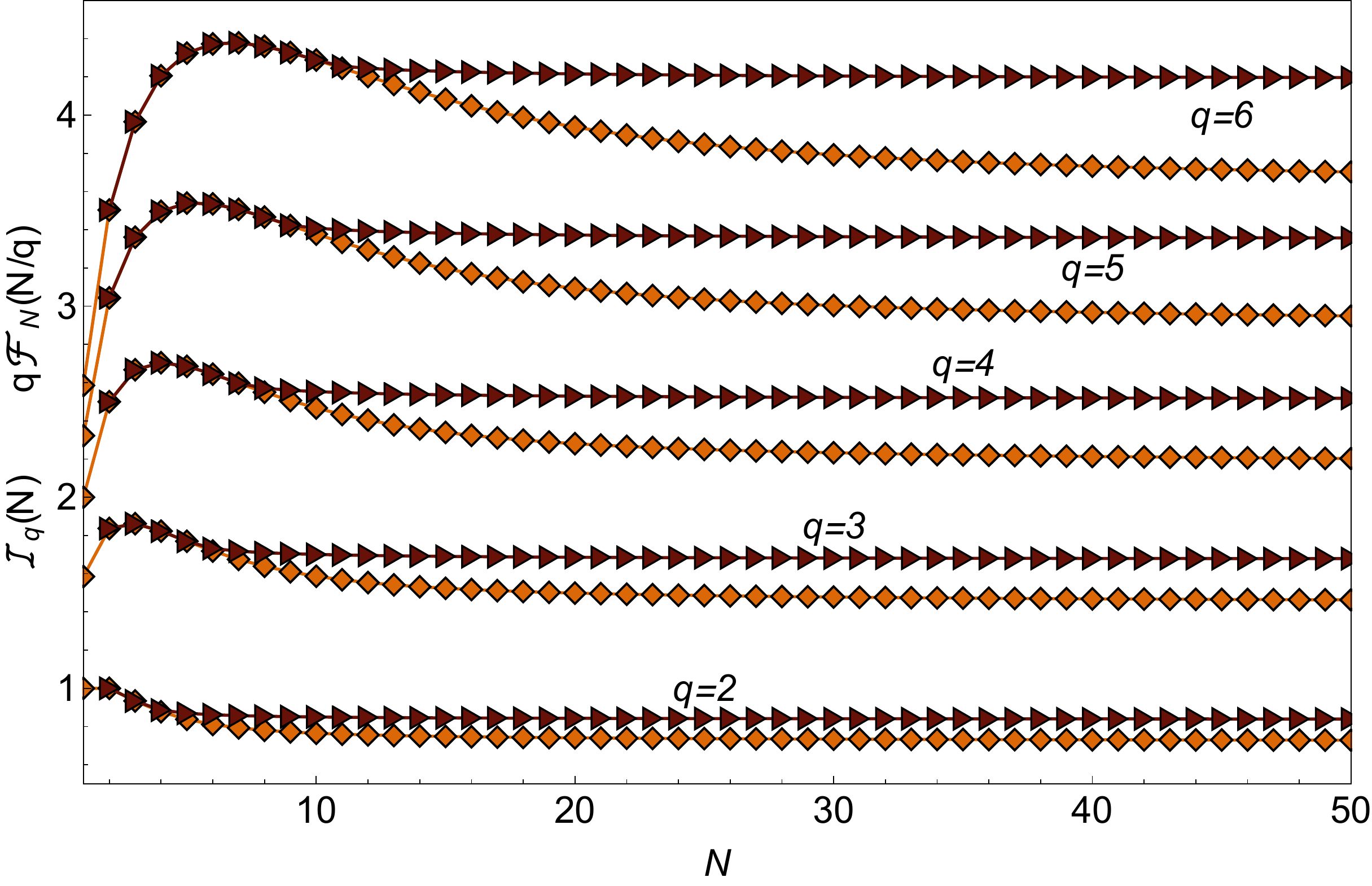}
\caption{Plot of the mutual information for the symmetric partition $q\mathcal{F}_N(N/q)$ (orange rhomboid markers) and the optimal mutual information $\mathcal{I}_q(N)$ (brown triangle markers) as a function of $N$, in units of nats. Top panel: Plots for a particular value on the number of partitions ($q=6$). As we can see, there exists a critical value $N^{\star}(q)$ (vertical solid red line) on the number of particles, below which the optimal partition is the symmetric one, while above it the optimal partition corresponds to an asymmetric partition. Moreover the optimal mutual information becomes maximal for a particular number of particles $\hat{N}(q)<N^{\star}(q)$ ($\hat{N}(s)$ is shown by the vertical solid blue line). Bottom panel: plot for various values of $q$, shown the same general features.}
\label{fig3}
\end{figure}

This, in turn, allows us to write down a general expression for $\hat{\mathcal{I}}_{q}(N) $, namely:
\beeq{
\hat{\mathcal{I}}_q (N)&=\mathcal{F}_N (n^+) + (q-1) \mathcal{F}_N (n^{-})\,,
 \label{eq:final}
}
with the constraint $n^{+}+(q-1)n^{-}=N$. Expression \eqref{eq:final} also contains the symmetric solution, corresponding to $n^{+}=n^{-}=N/q$.
Thus, the optimal value of the mutual information can always be written as 
\beeq{
\mathcal{I}_q(N)=\max_{0\leq n^{+}\leq N}\left[\mathcal{F}_N(n^+)+(q-1)\mathcal{F}_N\left(\frac{N-n^+}{q-1}\right)\right]\,,
\label{eq:f}
}
and is plotted in Fig. \ref{fig3} (brown triangle markers), and compared to the mutual information achieved for the symmetric partition, corresponding to $n_i=N/q$  for $i=1,\ldots,q$ (orange rhomboid markers).
We see that, for a fixed $q$, there exists a critical value of particles $N^\star(q)$ such that for $N\leq N^\star(q)$ the optimal partition is always the symmetric one, while for $N>N^\star(q)$, the optimal partition is asymmetric. We also found that for $q\gg 1$, this critical number of particles is given by $N^*(q)\simeq 2.1803 q$. Additionally, Fig. \ref{fig3} shows that there exists a value $\hat{N}(q)<N^\star(q)$ for which $\mathcal{I}_q(N)$ is maximal at fixed $q$. 

We obtain an asymptotic value of $\hat{N}(q)$ by noticing that, since this maximum is achieved for a symmetric partition we must have that $\hat{N}=nq$. The optimal mutual information is then given by $\hat{N} I[X,\bm{K}]= q\mathcal{F}_{nq}(n)$, so that for a large number of partitions, and using Eq. (\ref{eq:3b}) we obtain $\hat{N}(q)\sim 1.338 q$. Note also that as $N$ goes to infinity, the value of $\mathcal{I}_q(N)$ given by Eq. \eqref{eq:f} will be dominated by the $q-1$ smaller partitions, yielding the asymptotic value of $\mathcal{I}_q(N)\sim (q-1)0.8371$, as shown in the top panel of Fig. \ref{fig3}.\\

\section{Conclusions}
\label{sec6}
In a \LS\ engine that can use an ideal gas with $N$ particles, and for which the box can be partitioned into $q$ partitions with walls that can move to extract work via a quasi-static process, the average extracted work is proportional to the information retained in the counts of how many particles fall into each partition about the single particle locations. The cost of running a memory that contains counts of how many particles are in each partition is proportional to the information retained about the ensemble locations. Run cyclically, the engine's efficiency is thus limited by the difference---the information retained in memory that is not relevant with respect to work extraction. This provides 
a non-negative lower bound on engine dissipation.

We calculated the maximum average extracted work, by optimizing over the initial wall locations, for given $N$ and $q$. For fixed $q$, there is a critical value $N^{\star}(q)$ below which the optimal partitioning is symmetric (partitions of equal volume), and above which an asymmetric solution is preferable. The maximum value of the average extractable work occurs at $\hat{N}(q)<N^{\star}(q)$. Asymptotically, as $N\rightarrow \infty$, $\hat{N}(q)$ is linear in $q$, as is the maximal extractable average work.

The extension of \LS's engine we have explored here is basic in the sense that it allows for work extraction only via movement of the partitions along the $x$-axis, without building in additional work extraction mechanisms, such as used e.g. in \cite{Horowitz_2011}, and in that it uses a naive, counting observer, without requiring an optimal observer, as discussed in \cite{CB}.

\appendix

\section{Notation and Calculations}
\label{A}
Vectors are bold face symbols, e.g., $\bm{x} = (x_i, \dots, x_N)$. Entropy is written as $H[X] = -\langle \ln{p(x)} \rangle_{p(x)}$, and $H[\bm{X}] = -\langle \ln{p(\bm{x})} \rangle_{p(\bm{x})} = -\langle \ln{p(x_i, \dots, x_N)} \rangle_{p(x_i, \dots, x_N)}$, where $\langle \cdot \rangle_{p}$ denotes the average over the probability distribution $p$. In the main text, we sometimes use capital letters to denote probability distributions, in this Appendix we use small letters. 

Conditional entropy is written as 
\begin{eqnarray}
H[X|Y] &=& - \langle \ln{p(x|y)} \rangle_{p(x,y)},\\
H[\bm{X}|Y] &=& -\langle \ln{p(\bm{x}|y)} \rangle_{p(\bm{x},y)} \\
&=& -\langle \ln{p(x_1, \dots, x_N|y)} \rangle_{p(x_1, \dots, x_N,y)},\\
H[X|\bm{Y}] &=& -\langle \ln{p(x|\bm{y})} \rangle_{p(x,\bm{y})} \\
&=& -\langle \ln{p(x|y_1, \dots, y_M)} \rangle_{p(x,y_1, \dots, y_M)},\\
H[\bm{X}|\bm{Y}] &=& -\langle \ln{p(\bm{x}|\bm{y})} \rangle_{p(\bm{x},\bm{y})} \\
&& \!\!\!\!\!\!\!\!\!\!\!\!\!\!\!\!\!\!\!\!\!\!\!\!\!\!\!\!\!\!\!\!\!\!\!\!\!\!\!\!\!\! = -\langle  \ln p(x_1, \dots, x_N|y_1, \dots, y_M) \rangle_{p(x_1, \dots, x_N,y_1, \dots, y_M)},
\end{eqnarray}
respectively, and mutual information is defined the usual way \cite{CoverThomas}, written as
\begin{eqnarray} 
I[X,Y] &=& H[X] - H[X|Y], \label{MIdef}
\end{eqnarray}
which can be written as the relative entropy between the joint distribution $p(x,y)$ and the product of the marginals, or between the conditional $p(x|y)$ and the marginal $p(x)$:
\begin{eqnarray}
I[X,Y] &=& \left\langle \ln{\left[ {p(x,y) \over p(x) p(y)} \right]}\right\rangle_{p(x,y)} = \left\langle \ln{\left[ {p(x|y) \over p(x)} \right]}\right\rangle_{p(x,y)}.
\end{eqnarray}
While mutual information is defined for two random variables, total correlation, also called multi information, is defined as the relative entropy between the joint distribution of $N$ random variables $p(X_1, \dots, X_N)$ and the product of the marginals, $\prod_{i=1}^N p(X_i)$. We write it as
\begin{eqnarray}
I[\bm{X}] &=& I[X_1, \dots , X_N] \\
&=& \left\langle \ln{\left[ {p(x_1, \dots, x_N) \over \prod_i^{N} p(x_i )} \right]} \right\rangle_{p(x_1, \dots, x_N )} \\
&=& \sum_{i=1}^N H[X_i] - H[X_1, \dots, X_N]
\end{eqnarray}
Conditional total correlation, also called conditional multi information, is defined similarly, with all distributions conditional on the outcome of another random variable, say $Y$:
\begin{eqnarray}
I[\bm{X},Y] &=& I[X_1, \dots , X_N | Y] \\
&=& \left\langle \ln{\left[ {p(x_1, \dots, x_N | y) \over \prod_i^{N} p(x_i | y)} \right]} \right\rangle_{p(x_1, \dots, x_N, y)}\\
&=& \sum_{i=1}^N H[X_i|Y] - H[X_1, \dots, X_N|Y]~.
\end{eqnarray}
Similarly for $\bm{y} = (y_i, \dots, y_M)$, we have 
\begin{eqnarray}
I[\bm{X},\bm{Y}] &=& H[\bm{X}] - H[\bm{X}|\bm{Y}]\\
&=& H[\bm{Y}] - H[\bm{Y}|\bm{X}]~.
\end{eqnarray}
Using these definitions, it is easy to see that
\begin{eqnarray}
\!\!\!\!\!\!\!\!\!\!\!\!\!\!\!\!\!\!&&I[\bm{X},  \bm{K}] - \sum_{i=1}^{N}I[X_i,  \bm{K}] \notag\\
\!\!\!\!\!\!\!\!\!\!\!\!\!\!\!\!\!\!&=& H[\bm{X}] - H[\bm{X} | \bm{K}] - \sum_{i=1}^N H[X_i] + \sum_{i=1}^N H[X_i|\bm{K}] \\
\!\!\!\!\!\!\!\!\!\!\!\!\!\!\!\!\!\!&=& I[\bm{X} | \bm{K}] - I[\bm{X}]~.
\end{eqnarray}
Since the particles are identical, we have 
\begin{equation}
NI[X,  \bm{K}] = \sum_{i=1}^{N}I[X_i,  \bm{K}]~,
\end{equation}
and because they are non-interacting, the variables $X_i$ are independently distributed before the measurement, i.e. 
$P(\bm{X})=\prod_{i=1}^N P({X_i})$, whereby $I[\bm{X}] = 0$.

Eq. (\ref{Iirrel-2}) in the main text follows directly from these observations: 
\begin{eqnarray}
I[\bm{X},  \bm{K}] - NI[X,  \bm{K}] = I[\bm{X} | \bm{K}]~.
\end{eqnarray}

\section{Validity of our results in the quasi-static limit} \label{sec:appendix}

In response to a request by one of the reviewers, this short appendix recalls elementary definitions in thermodynamics \cite{callen1998thermodynamics} that are used throughout this paper. Specifically, we assume: i) that the system is at all times in contact with a heat bath at constant temperature $T$, and ii) that the engine is operated quasi-statically during the complete work extraction process. Both of these assumptions are important because they allow us to model the engine as a set of coupled ideal gases instead of using a stochastic description for the collisions between the particles and the walls. This means that our results are not restricted to the thermodynamic limit, $N\to \infty$, but hold for any number of particles. 

Assumption i) implies that for long enough times, our system will be Markovian, while ii) guarantees that dynamical processes occur over long time scales. 

Wall $i$ separates two chambers, of lengths $\ell_i$ and $\ell_{i+1}$, respectively.  Each time a particle hits the $i$-th wall the latter changes its position by a small amount $\epsilon$, and thus, after a fixed time $t$, its displacement is given by $t (m_{i+1}-m_{i})$, where $m_i$ is the number of collisions impinging on the wall coming from the $i$-th partition. It is clear that $m_i$ is a random variable, but given that particles do not interact with each other, and given that the system has a constant temperature, the number of collisions from either side can be described as a Poission process with constant rate. Hence, $\langle m_i \rangle \propto t k_i/\ell_i$ and similarly its variance fulfills that  $\text{Var}[m_i] \propto tk_i/\ell_i$. By taking the $\epsilon \to 0 $ limit it is clear that the time needed for any wall to move by a finite amount diverges as $1/\epsilon$, thus recovering the quasi-static limit. Note also that in this regime, the fluctuations of the position of the walls vanish as $\sqrt{\epsilon}$, and thus can be safely neglected. Therefore, when the engine is operated within this regime, the equilibrium condition is simply determined by the equality of pressure in all the partitions, as we used in the main text, but no further assumption on the value of $N$ is needed.

\acknowledgments
J.S., S.S. and R.D.H.R. thank The Abdus Salam International Centre for Theoretical Physics (ICTP) for hospitality and support during a visit in 2016 when the first part of this work was carried out. J.S. was partially supported by ICTP through the OEA-AC-98.  S.S. is grateful for funding from the Foundational Questions Institute (together with the Fetzer Franklin Fund), Grant No. FQXi-RFP-1820. I.P.C. acknowledges financial support from the project UNAM-DGAPA-PAPIIT IA103417 and support from the London Mathematical Laboratory, where he is an external fellow.

\bibliographystyle{apsrev4-1}
\bibliography{SZE_small}

\end{document}